\documentclass[]{zakoproc}
\usepackage{graphicx}
\begin{document}

\title{Power of Turbulent Reconnection: Star Formation, Acceleration of Cosmic Rays, Heat Transfer, Flares and Gamma Ray Bursts}
\author{Alex Lazarian}
\institute{Department of Astronomy, University of Wisconsin-Madison, USA} \markboth{A. Lazarian}{Power of Turbulent Reconnection}

\maketitle

\begin{abstract}
Turbulence is ubiquitous in astrophysical fluids. Therefore it is necessary to study magnetic reconnection in
turbulent environments. The model of fast turbulent reconnection proposed in Lazarian \& Vishniac 1999 has been
successfully tested numerically and it suggests numerous astrophysical implications. Those include a radically
new possibility of removing magnetic field from collapsing clouds which we termed "reconnection diffusion", 
acceleration of cosmic rays within shrinking filaments of reconnected magnetic fields, flares of reconnection,
from solar flares to much stronger ones which can account for gamma ray bursts. In addition, the model
reveals a very intimate relation between magnetic reconnection and properties of strong turbulence,
explaining how turbulent eddies can transport heat in magnetized plasmas. This is a small fraction
the astrophysical implications of the quantitative insight into the fundamental process of magnetic 
reconnection in turbulent media.

\end{abstract}

\section{Introduction}

Astrophysical fluids are turbulent. For instance, the interstellar medium (ISM) is known to be turbulent on scales ranging from AUs to kpc (see Armstrong et al 1995, Elmegreen \& Scalo 2004, Lazarian 2009, Chepurnov \& Lazarian 2010).Magnetized astrophysical plasmas generally have very large Reynolds numbers due to the large length scales involved and the fact that the motions of charged particles in the direction perpendicular to magnetic fields are constrained. Laminar plasma flows at these high Reynolds numbers are prey to numerous linear and finite-amplitude instabilities, from which turbulent motions readily develop. Indeed, observations show that turbulence is ubiquitous in all astrophysical plasmas (see also Leamon et al. (1998), Bale et al. (2005) for solar wind, Padoan et al. (2008) for molecular clouds and Schucker et al. (2004)), Vogt \& En§lin (2005) for the intracluster medium. The plasma turbulence is sometimes driven by an external energy source, such as supernova in the ISM (Norman \& Ferrara 1996), merger events and active galactic nuclei outflows in the intercluster medium (ICM) (Ensslin \& Vogt 2006), and baroclinic forcing behind shock waves in interstellar clouds. In other cases, the turbulence is spontaneous, with available energy released by a rich array of instabilities, such as MRI in accretion disks (Balbus \& Hawley, 1998), kink instability of twisted flux tubes in the solar corona (Galsgaard \& Nordlund 1997a, Gerrard \& Hood 2003), etc. Whatever its origin, the signatures of plasma turbulence are seen throughout the universe.

The textbook treatment of magnetic reconnection, i.e. the ability of magnetic field lines to change magnetic topology, ignores pre-existing
turbulence. This is a serious oversight as far as astrophysical fluids are concerned. Turbulence radically changes many processes, e.g. diffusion, cosmic
ray transport, advection of heat, and it would be strange if magnetic reconnection were not affected. Indeed, the model proposed in Lazarian \& Vishniac (1999, henceforth LV99) identified the way of how magnetic reconnection get enhanced by turbulence and quantified the expected reconnection rates. These predictions 
have been successfully tested in Kowal et al. (2009), which opened avenues for studies of the implications of magnetic reconnection in astrophysics.

In what follows we discuss the process of turbulent reconnection in section 2, discuss how turbulent reconnection
changes the star formation paradigm in section 3, outline the consequences of the LV99 process for cosmic
ray acceleration in section 4, explain why the turbulent reconnection is important for heat transfer in magnetized
plasmas in section 5. Sections 6 deals with the bursts of reconnection predicted by LV99 model and discusses
their relation to both solar flares and gamma ray bursts. Section 7 outlines the prospects of the research.

\section{3D Turbulent Reconnection}

LV99 proposed a model of fast reconnection in the presence of sub-AlfvŽnic turbulence in magnetized plasmas (see Fig. 1). They identified stochastic wandering of magnetic field-lines as the most critical property of MHD turbulence which permits fast reconnection.  As illustrated below, this line-wandering widens the outflow region and alleviates the controlling constraint of mass conservation. 

\begin{figure*}[t]
\vspace*{2mm}
\begin{minipage}[t]{6.7cm}
\begin{center}
\includegraphics[bb = 47 47 522 317,width=6.9cm,clip=]{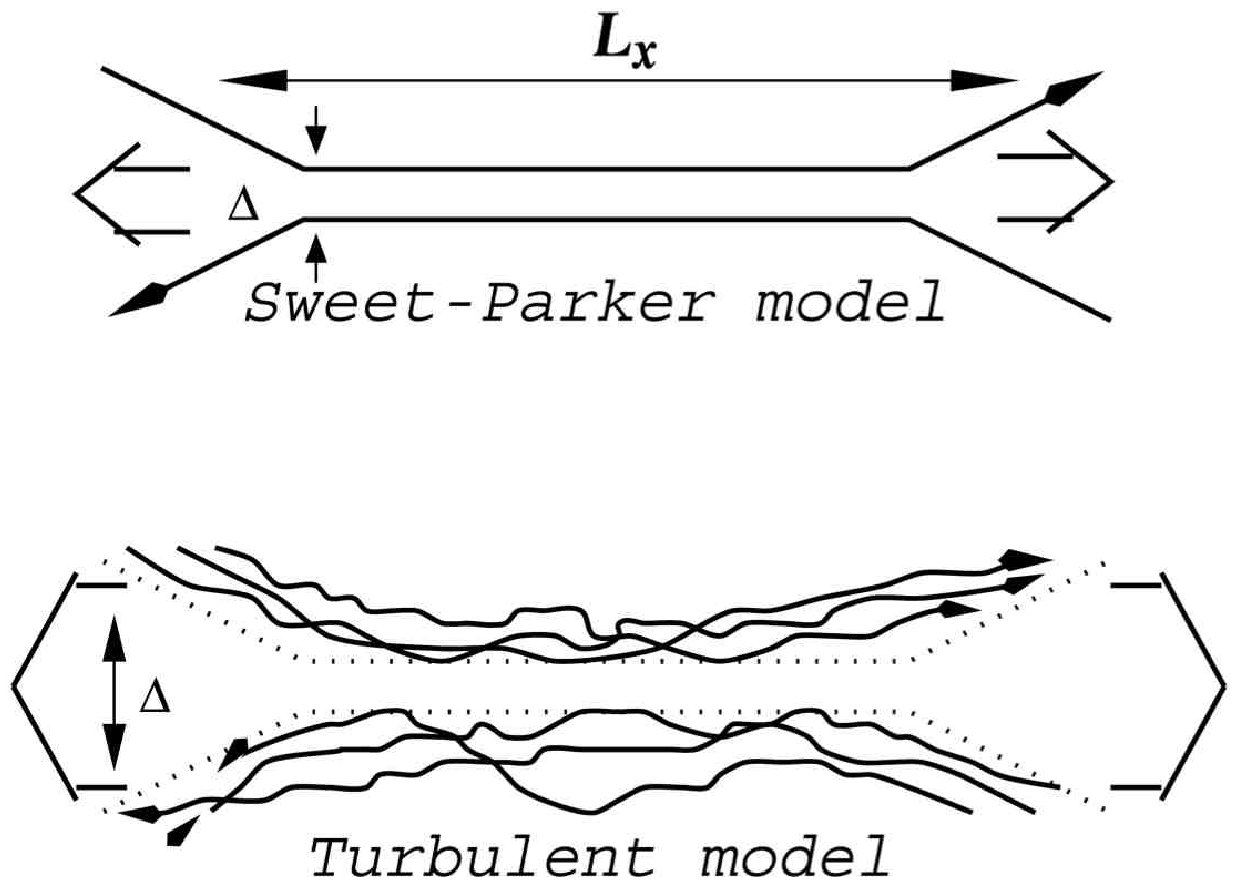}
\caption{Upper plot: Sweet-Parker model of reconnection. The outflow is limited by a thin slot Æ, which is determined by Ohmic diffusivity. The other scale is an astrophysical scale $L_x\gg \Delta$. Lower plot: Reconnection of weakly stochastic magnetic field according to LV99. The Goldreich-Sridhar (1995) model of MHD turbulence is used to account for the stochasticity of magnetic field lines. The outflow in the LV99 theory is limited by the diffusion of magnetic field lines, which depends on field line stochasticity. From Lazarian et al. 2004.} 
\end{center}
\end{minipage}\hfill
\begin{minipage}[t]{6.7cm}
\begin{center}
\includegraphics[bb = 47 47 522 317,width=8 cm,clip=]{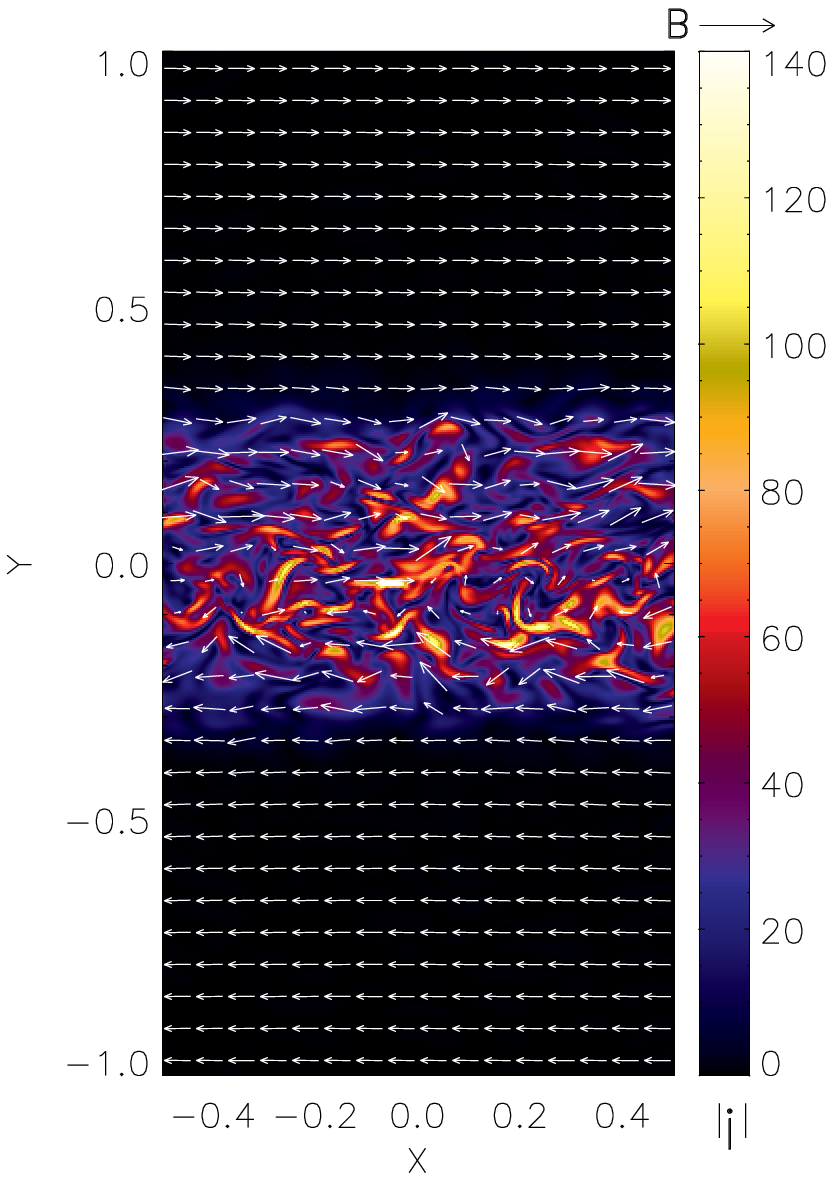}
\caption{Current intensity and magnetic field configuration during stochastic reconnection.  We show a slice through the middle of the computational box in the xy plane after twelve dynamical times for a typical run.  The shared component of the field is perpendicular to the page.  The intensity and direction of the magnetic field is represented by the length and direction of the arrows.  The color bar gives the intensity of the current.  From Kowal et al. 2009.} 
\end{center}
\end{minipage}
\end{figure*}

The LV99 model is radically different from its predecessors which also appealed to the effects of turbulence. For instance, unlike Speiser (1970) and Jacobson \& Moses (1984) the model does not appeal to changes of microscopic properties of plasma.  The nearest progenitor to LV99 was the work of Matthaeus \& Lamkin (1985) Matthaeus \& Lamkin (1986), who studied the problem numerically in 2D MHD and who suggested that magnetic reconnection may be fast due to a number of turbulence effects, e.g. multiple X points and turbulent EMF. However, these papers did not address the important role of magnetic field-line wandering,  and did not obtain a quantitative prediction for the reconnection rate, as did LV99.

\begin{figure*}[t]
\vspace*{2mm}
\begin{minipage}[t]{6.7cm}
\begin{center}
\includegraphics[bb = 47 47 522 317,width=6.7cm,clip=]{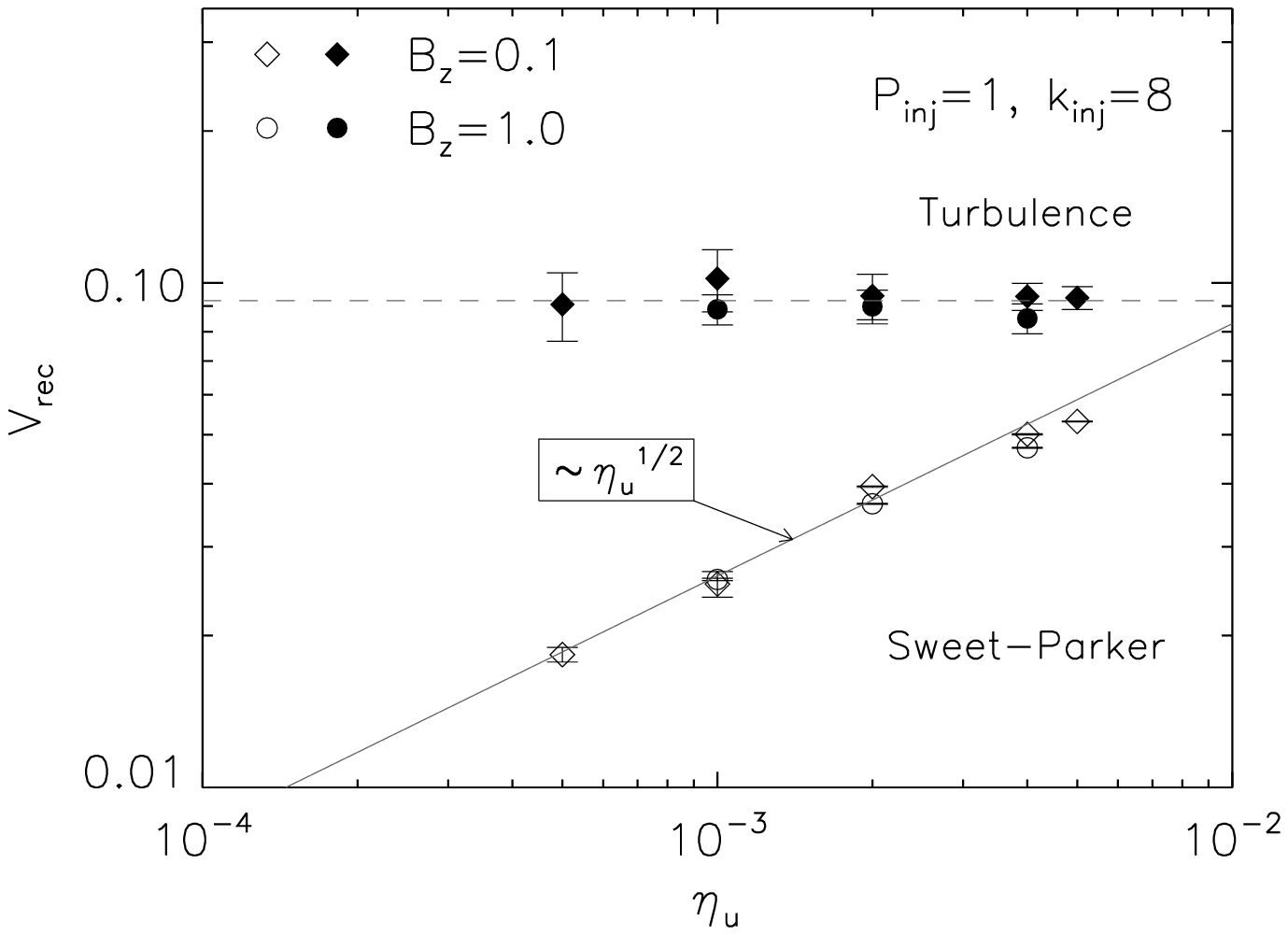}
\caption{Reconnection speed versus resistivity. Laminar corresponds to hollow symbols, and the stochastic reconnection to the filled symbols. The symbol sizes indicate the uncertainty in the average reconnection speeds and the error bars indicate the variance. From Kowal et al. 2009.} 
\end{center}
\end{minipage}\hfill
\begin{minipage}[t]{6.7cm}
\begin{center}
\includegraphics[bb = 47 47 522 317,width=6.7cm,clip=]{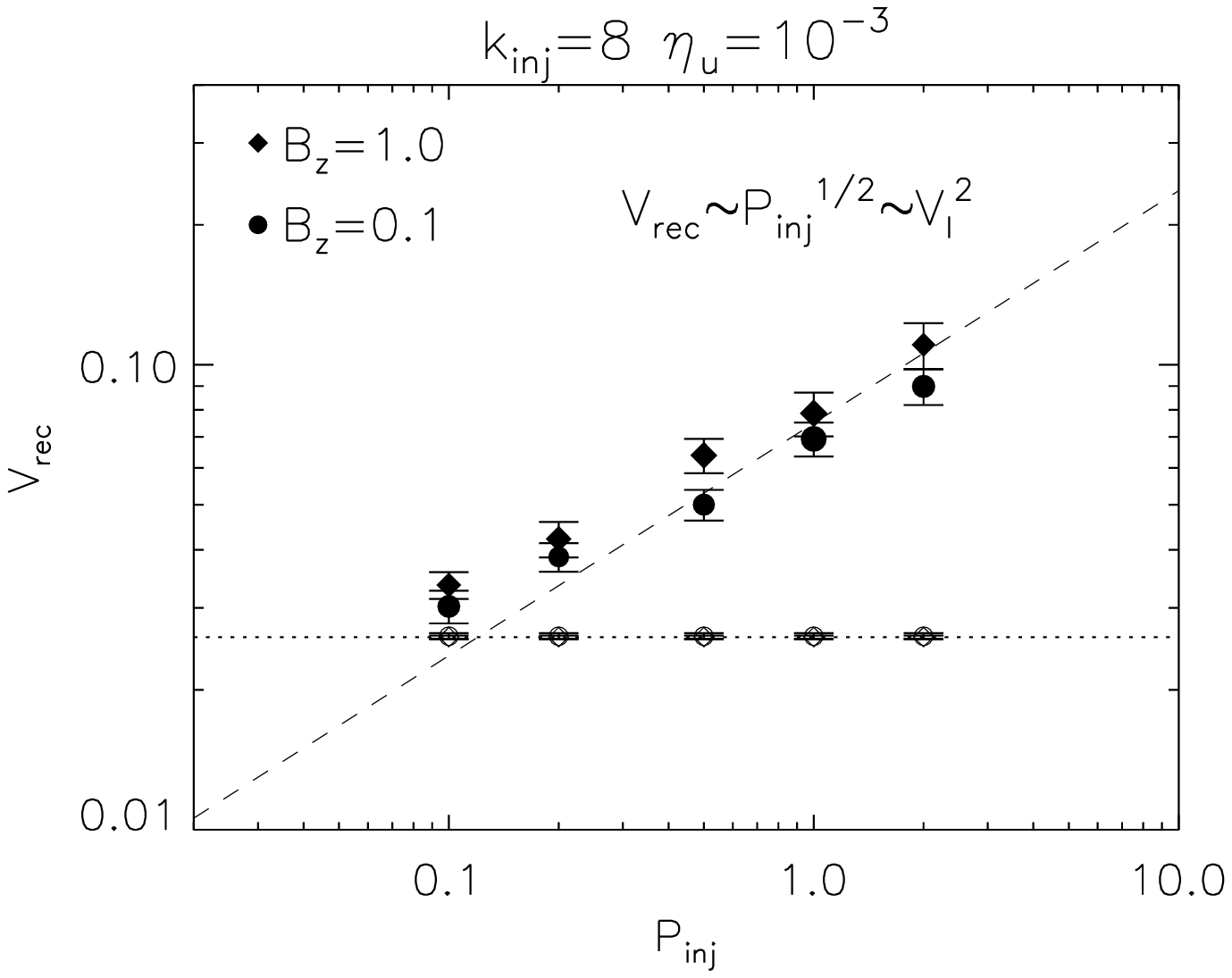}
\caption{Reconnection speed versus input power for the driven turbulence. The dashed line is a fit to the data with the predicted in LV99 dependence of  $P_{inj}^{1/2}$. From Kowal et al. 2009.} 
\end{center}
\end{minipage}
\end{figure*}

The LV99 model has been successfully tested recently in Kowal et al. (2009) (see also higher resolution results in Lazarian et al. (2010)).  The simulations are periodic in the direction of the shared field (the $z$ axis) and are open in the reversed direction (the $x$ axis).  The external plasma pressure is uniform and the magnetic fields at the top and bottom of the box are taken to be the specified external fields plus small perturbations to allow for outgoing waves. The driven turbulence mimics turbulence pre-existing in most astrophysical environments (see Fig. 2). 
Figure 3 and 4 illustrates some of the results obtained with the simulations. First of all, we see no dependence of the reconnection rate on resistivity, while the dependence of the reconnection rate on the turbulence input power corresponds well to the predictions of the LV99 model.

The success of LV99 in identifying an MHD turbulence mechanism for fast reconnection leads to a conflict with certain conventional beliefs. As the predicted reconnection velocity is independent of magnetic diffusivity $\eta$ the LV99 theory implies that field-line topology should change in MHD plasmas at a finite rate even in the limit of infinite Lundquist number. This contradicts the accepted wisdom that magnetic field-lines should be ÒnearlyÓ frozen-in to very high-conductivity MHD plasmas. It is implicit in the LV99 theory that the standard Alfven Theorem on magnetic-flux conservation must be violated for $\eta\rightarrow 0$ (Vishniac \& Lazarian, 1999). LV99 and later more formal mathematical studies (see Eyink 2011) showed  that "flux freezing" is incompatible with the description of magnetic fields in turbulent conducting fluids. Eyink, Lazarian \& Vishniac (2011, henceforth ELV11) established the equivalence of the LV99 and recent mathematical results. They provided a new derivation of LV99 predictions appealing to a well-studied property of turbulent fluids: the Richarson diffusion (see also Lazarian, Eyink \& Vishniac 2011).  

We would like to stress that LV99 model is not in conflict with the studies of magnetic reconnection in collisionless plasmas that have been a major thrust of the plasma physics community (see Shay et al. 1998, Daughton et al. 2006). Unlike latter studies, LV99 deals with turbulent environments. It shows that local reconnection rates are influenced by plasma effects, e.g. kinetic effects of Hall effects, but the overall or global reconnection rate is determined by the turbulent broadening of the reconnection region. This conclusion was confirmed in numerical simulations by Kowal et al. (2009, henceforth KX09). We note the complementary nature of the LV99 model and more recent studies of tearing instability of laminar reconnection layers (see Loureiro et al. 2011). The latter research clarifies how magnetic reconnection may proceed in special (from the ISM perspective) environments where the initial state of conducting fluid is laminar. 

\section{Reconnection and Changing Paradigm of Star Formation}

Star formation presents an important avenue for applying the LV99 theory. The theory was formulated for both collisional and collisionless turbulent plasmas (see the quantitative elaboration of the latter point in ELV11) and was extended to the partially ionized gas in Lazarian, Vishniac \& Cho (2004). Therefore it can be applied to molecular clouds. In fact, Lazarian (2005)  identified fast magnetic reconnection in turbulent fluid as a promising way of changing mass loading of magnetic flux during star formation. The corresponding process was termed "reconnection diffusion" (Lazarian et al. 2010). 

The reconnection of flux tubes which takes place in turbulent media as shown in Fig. 5. The mixing is happening as new magnetic flux tubes are constantly formed from the magnetic flux tubes that belong to different eddies. In the figure two adjacent eddies are 
shown and the process is limited to the effects of eddies of a single scale. It is clear that 
plasmas which was originally entrained over different flux tubes gets into contact along the new
emerging flux tubes. The process similar to the depicted one takes place at different scales down
to the scale of the smallest eddies. Molecular diffusivity then takes over.

The process of reconnection diffusion is due to eddies that are perpendicular to the {\it local} direction of
magnetic field. This direction, in general, does not coincide with the mean magnetic field direction. Therefore in the lab system of reference related to the mean magnetic field the diffusion of magnetic field and plasmas will happen both parallel and perpendicular to the mean magnetic field direction.

Before this development, it had been universally believed that to change the mass loading on magnetic field lines one should
invoke imperfect coupling of ions and neutrals, i.e. the process which is referred in astrophysics as "ambipolar diffusion". The ambipolar
diffusion was the corner stone of the old star formation paradigm. It, however, is facing problems explaining some of the observations (see Crutcher 2012). 

\begin{figure}
\centering
  \includegraphics[height=.20\textheight]{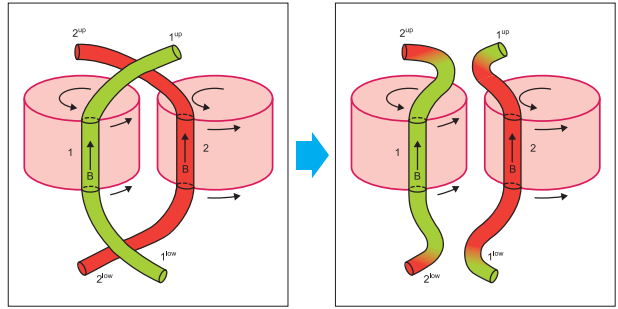}
  \caption{Reconnection diffusion. llustration of the mixing of matter due to reconnection as two flux tubes of different magnetic field strength interact. From Lazarian 2011a.}
\end{figure}

Reconnection diffusion successfully explains observational results which are puzzling within the old paradigm. For instance, Troland \& Heiles (1986) showed that no correlation exists between magnetic field and density in diffuse interstellar gas. Such a correlation is expected if one considers compressions of the magnetic field and gas of relatively high ionization for which the ambipolar diffusion is negligible. Reconnection diffusion, on the contrary, is expected to increase entropy by removing the correlations. This conclusion was confirmed by our simulations  (Santos-Lima et al. 2010, henceforth SX10) that show that the diffusion is fast without ambipolar diffusion (cp. Heitch et al. 2004), supporting the intuitive notion that turbulent transport processes are independent of microphysic diffusivity. 

We can reliably interpret our results on reconnection diffusion as LV99 (see also KX09) showed that turbulent reconnection is independent of the nature of resistivity, i.e. the same for Ohmic, numerical or anomalous (due to Hall or other plasma effects) resistivities. 	
In the presence of gravity one expects that diffusive processes would allow heavy matter, i.e. gas, to flow to the center of gravity and light matter, i.e. magnetic field, to diffuse away. In SX we reported the removal of magnetic field for various initial conditions, including those emulating collapsing supercritical and subcritical clouds. The rates of the magnetic field removal were found to be consistent with our predictions of turbulent magnetic diffusivity in Lazarian (2006). Further simulations (Santos-Lima et al. 2011) demonstrated the efficient removal of the magnetic field from circumstellar disks in the process of their formation from turbulent magnetized gas. This explains the otherwise puzzling results in Shu et al. (2006) on rapid removal of magnetic fields from cores and disks.  We found that the properties of circumstellar disks obtained in simulations with a turbulent initial state of matter correspond to the observed ones.

We believe that the reconnection diffusion presents the last missing piece for constructing the new paradigm of star formation where turbulence and turbulent feedback play the central role. In Lazarian (2011a) we claim that the acid test for the theory presented by results of Crutcher et al. (2009, 2010, henceforth CX) who presented evidence against the traditional theory on the basis of their Zeeman measurements in the cores and the envelopes of a molecular cloud. The ambipolar diffusion paradigm requires the mass to flux ratio to be smaller at the core and larger at the envelope. CX reported the opposite picture. 

In is easy to observe that, qualitatively, reconnection diffusion is consistent with the CX results. Indeed, observationally, it is known that the level of turbulence drops in the cores. Reconnection diffusion slows down with the decrease of turbulent velocity (Lazarian 2006, SX10), so we expect a slower transport of magnetic flux out of the core as compared to the envelope. This is expected to lead to a larger mass to flux ratio in the core compared to the envelope, in agreement with observations (Lazarian 2011a).

\section{Acceleration of Energetic Particles}

Turbulent magnetic reconnection described in LV99 envisages the existence of shrinking loops of 3D magnetic field. The energetic
particles entrained on such a loop are expected to accelerate. It is obvious from the Fig.~6 that the energetic particle on a shrinking loop
will get the energy all the time as the loop shrinks. This can be seen as the consequence of the preservation of the phase volume in 
the absence of collisions. Another way to explain why the acceleration is the first order Fermi is to consider energetic particles bouncing
back and forth between converging mirrors of not reconnected flux\footnote{In the presence of fast scattering that preserves isotropy of
energetic particle distribution the nature of the acceleration changes, as particles equally efficient in scattering from the divergent flows
presented by the outflowing matter. However, in real astrophysical situations the energetic particles are entrained with the magnetic field.}. 

The corresponding calculations of the acceleration were performed in de Gouveia dal Pino \& Lazarian (2005, henceforth GL05) (see also Lazarian 2005).
The backreaction of the accelerated particles on magnetic field has not been studied in 3D yet. However, Drake et al. (2006, henceforth DX06) provided
a possible model of the backreaction considering 2D closed loops within his preferred model of collisionless reconnection. 

Since then, the acceleration of energetic particles has been invoked to explain the origin of the anomalous cosmic rays explored by 
Voyagers (Lazarian \& Opher 2009, Drake et al. 2010), the origin of the cosmic ray anisotropy observed by MILAGRO and ICECUBE
in the direction of the heliospheric magnetotail (Lazarian \& Desiati 2010). Naturally, this is just a start of the exploration of the 
consequences of the new first order Fermi acceleration mechanism.  

\begin{figure*}[t]
\vspace*{2mm}
\begin{minipage}[t]{6.7cm}
\begin{center}
\includegraphics[bb = 47 47 522 317,width=6.7cm,clip=]{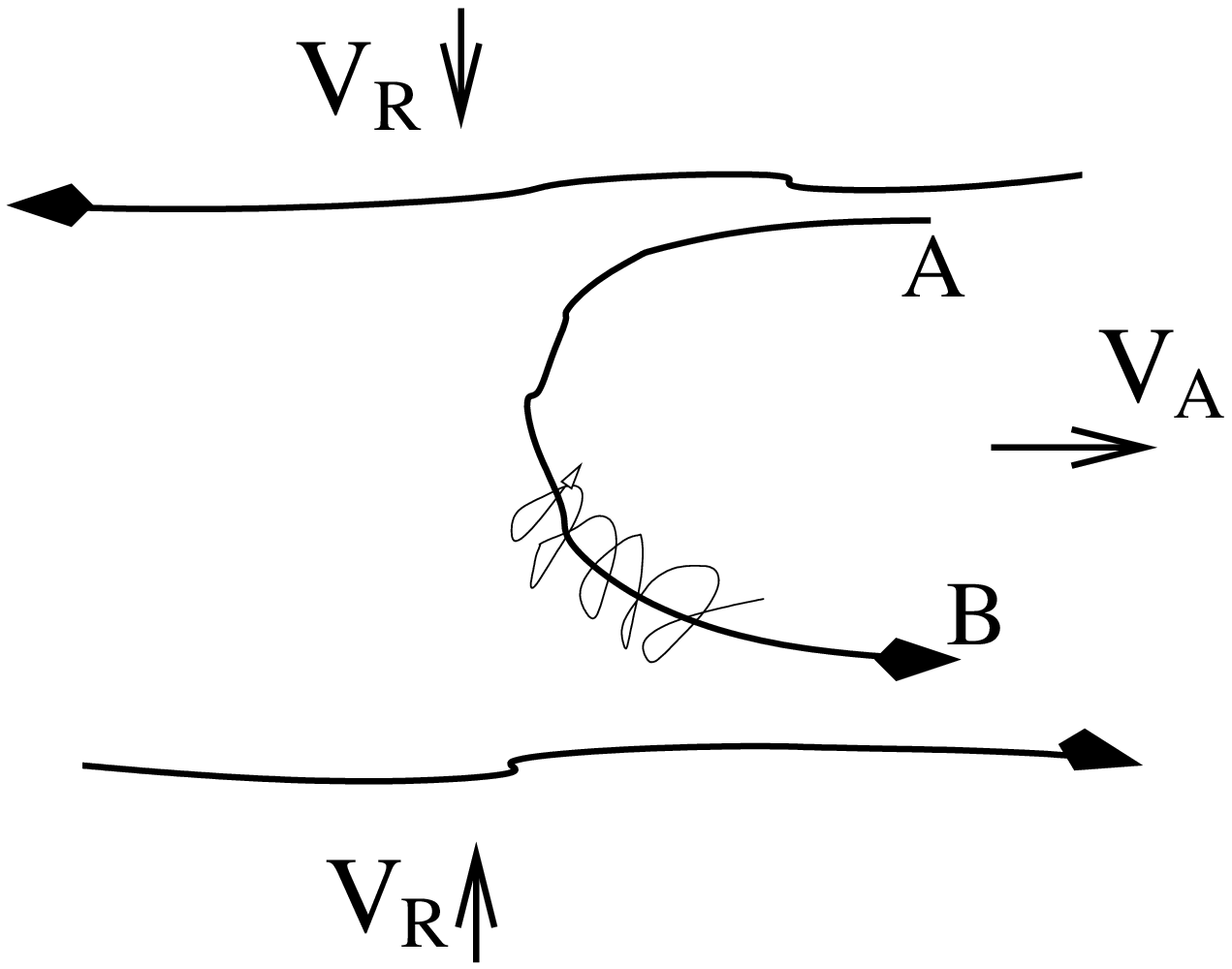}
\caption{Cosmic rays spiral about a reconnected magnetic
field line and bounce back at points A and B and gains energy. The reconnected
regions move towards each other with the reconnection velocity
$V_R$ of the order of $V_A$.  From Lazarian 2005.} 
\end{center}
\end{minipage}\hfill
\begin{minipage}[t]{6.7cm}
\begin{center}
\includegraphics[width=6.3cm,clip=]{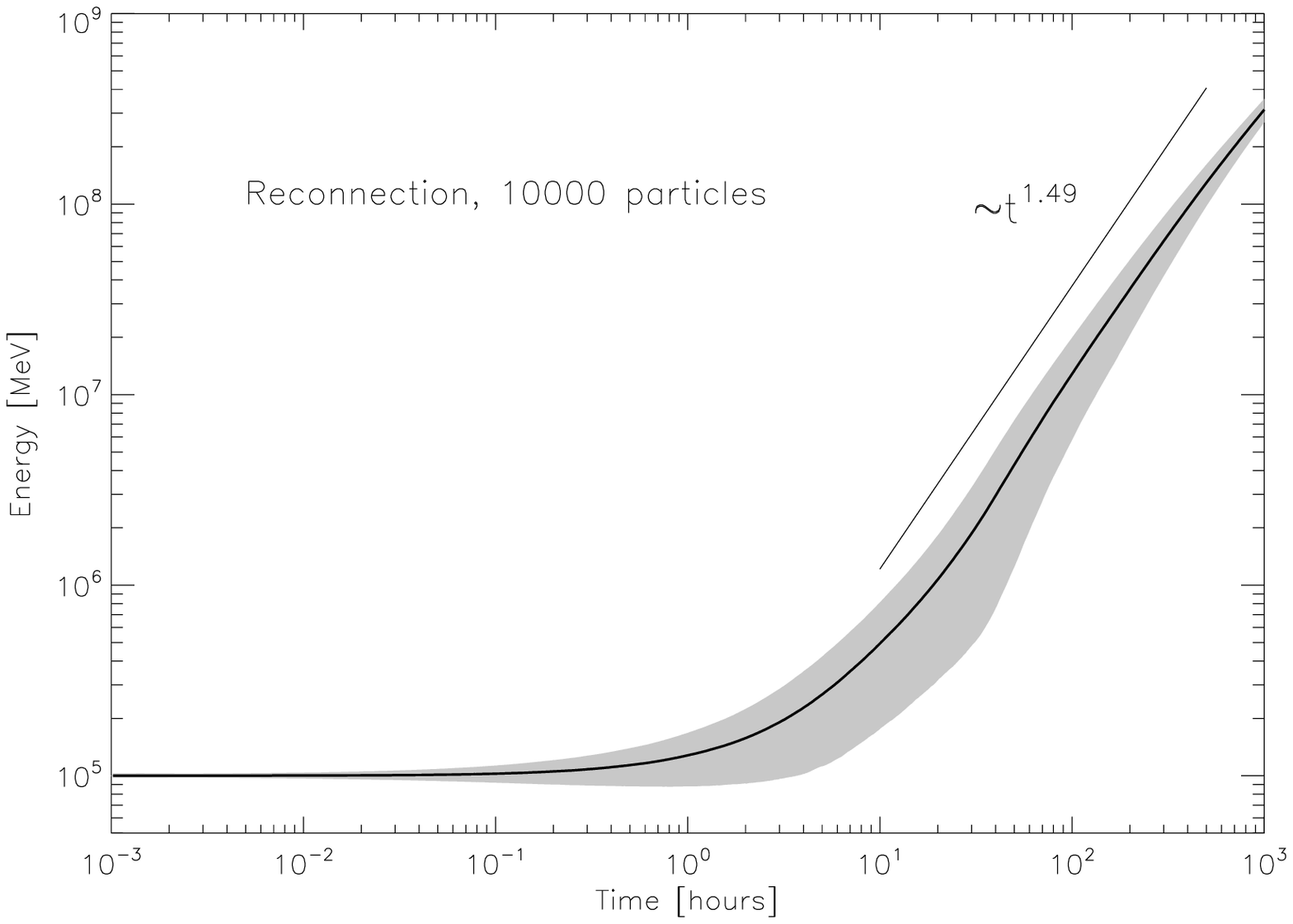}
\caption{Acceleration of particles in turbulent reconnection layers. Particle energy evolution averaged over 10,000 particles with the initial energy $E_0=10^5$ MeV and random initial positions and directions. From Lazarian et al. 2010} 
\end{center}
\end{minipage}
\end{figure*}

Recent numerical studies in Kowal et al. (2011a) showed that the acceleration in 3D and 2D are rather different. Therefore, while on
the surface the processes discussed in GL05 and DX06 are very similar, the astrophysically relevant studies should be performed not in 2D,
as in DX06, but in 3D. The efficiency of the acceleration in turbulent reconnection has been confirmed in Kowal et al. (2011b). Fig. 7 illustrates some recent numerical simulations of 3D acceleration.

\section{Heat Transfer in Magnetized Plasmas}

The reconnection diffusion concept that we discussed above is not limited to star formation. In the textbook
explanation of heat transfer in plasmas it is assumed that
magnetic field lines always preserve their identify in highly conductive plasmas even in turbulent flows. In this
situation the diffusion of charged particles perpendicular to magnetic field lines is very restricted. Thus
it is accepted that the mass loading of magnetic field lines does not change and density and magnetic field compressions follow each other. All these assumptions are violated in the presence of reconnection diffusion.

For heat transfer in magnetized plasma two processes are important (see Lazarain 2006). The first one, field wondering, described for the first time for the Goldreich \& Sridhar (1995) model of turbulence in LV99. The deviations of the magnetic field its mean direction allow the electrons (which are the fastest particles in plasma)
to diffuse perpendicular to the mean magnetic field. The second process is related to the turbulent advection of heat by the eddies of magnetized plasmas. The latter is clearly related to the reconnection diffusion and the fast reconnection of turbulent magnetic field (see ELV11). Numerically the efficiency of heat transfer by subAlfvenic turbulence
was described in Cho et al. (2003) with the justification of the relation of the low Lundquist number simulations
and high Lundquist number astrophysical turbulence provided by LV99. It is the independence of the
reconnection on the Lundquist number predicted by LV99 that allows the use of numerical simulations to explore
both the case of star formation and heat transfer.

Interestingly enough, the field wondering is also closely related to fast reconnection. LV99 demonstrated that fast
reconnection of the fields of different eddies makes the GS95 model self-consistent. The analysis of the deep
connection of the fast reconnection and the properties of Alfvenic turbulence is provided in ELV11.

The relative importance of the two processes was discussed in Lazarain (2006). Fig. 8 shows both the parameter
space where the regions of one process dominating the other are identified. The subpanel also shows that
for actual clusters of galaxies the heat advection by turbulent motions is the most efficient process.

\begin{figure}
\centering
  \includegraphics[height=.25\textheight]{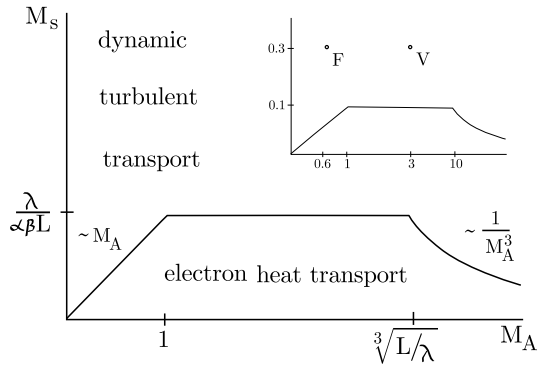}
  \caption{Parameter space for particle diffusion or turbulent diffusion to
dominate: application to heat transfer. Sonic Mach number $M_s$ is ploted
against the Alfven Mach number $M_A$. The heat transport is dominated by the
dynamics of turbulent eddies is above the curve (area denoted "dynamic turbulent
transport") and by thermal conductivity of
 electrons is below the curve (area denoted "electron heat transport"). Here $\lambda$ is the mean free path of the
electron, $L$ is the driving scale, and $\alpha=(m_e/m_p)^{1/2}$, $\beta\approx 4$. {\it Example of theory application}: The panel in the right upper corner of the figure illustrates
heat transport for the parameters for a cool
core Hydra cluster (point ``F''), ``V'' corresponds
to the illustrative model of a cluster core in Ensslin et al. (2005).  Relevant parameters were used 
for $L$ and $\lambda$. From
Lazarian (2006).}

\end{figure}

\section{Solar Flares and Gamma ray bursts}

The rate of reconnection in LV99 model depends on the level of turbulence. This provides a natural
explanation for solar flares and other flaring explosive phenomena. To get a solar flare one should
have first a period of the accumulation of magnetic flux of different polarity, i.e. the period when the
reconnection velocity is small. When the level of turbulence is low, LV99 predicts low the reconnection rates. 
The situation of slow reconnection is unstable, however. The reconnection induces turbulence, e.g. 
through the outflow, and this results in higher reconnection rates. The expected process is of self-accelerating, 
explosive nature. One can say that we are dealing with the {\it reconnection instability}.

For solar reconnection several of the predictions of the LV99 model have been tested. For instance, Ciaravella \& Raymond (2008) successfully tested the LV99 prediction of reconnection layers being thick and not X-point type.
Another prediction in LV99 was that reconnection events can stimulate neighboring regions to reconnect. This
effect was observed in Sych et al. (2009). 

ELV11 explain why the reconnection would follow LV99 even for collisionless plasmas. Therefore we expect
turbulent reconnection to dominate in both collisional (e.g. the ISM) and collisionless environments. Strongly
magnetized ones present the best case for bursts of reconnection. For instance, in Lazarian et al. (2003)
it was argued that gamma ray bursts can arise from LV99 type reconnection. This idea was further elaborated
and got observational support in a high impact paper by Zhang \& Yan (2011).

\section{Other Implications and Prospects}

The list of the possible implications of the turbulent reconnection is not limited by the examples above.
Turbulence is ubiquitous in astrophysical environments and therefore the turbulent reconnection is
also ubiquitous. The fact that LV99 provides solid analytical predictions of the reconnection rates should
help to parameterizing the reconnection and its effects in numerical codes.

Let us present a couple of implications which have not been discussed yet in detail in the literature. We discussed the acceleration of energetic particles in reconnection events. Similarly to energetic particles,
charged grains can be accelerated during magnetic reconnection. This process may be, for instance, important
for accelerating dust in accretion disks. Dust relative velocities control the rates of dust shattering and coagulation,
thus determining the dust size distribution which is important for chemical reactions, light propagation etc.  The dust grain velocities also influence absorption of metals, thickness of dusty disks and mixing gas with dust. 

In addition, while we discussed the heat transport, a similar process is important in redistribution of metals in the
ISM. Reconnection diffusion is a very basic process which can provide efficient mixing of elements through the galactic disk. 

Our studies of the implications of turbulent magnetic reconnection are at its infancy. We still have to clarify
many processes. For instance, it was argued in Lazarian \& Desiati (2010) that reconnection being an
intrinsic part of turbulent cascade should induce the first order Fermi acceleration even in the case
of pure MHD turbulence. Some of the results in Kowal et al. (2011) can be interpreted as the detection
of the predicted effect, but this is not conclusive. More studies are clearly required. 

{\bf Acknowledgment}. The research is supported by the Center for Magnetic Self-Organization (CMSO). I would like to thank Professor Alvaro Ferraz (IIP-UFRN-BRASIL) for his hospitality at IIP where part of this work was carried out. I also ackowledges MCTI  BRASIL  for their financial support during my stay at IIP-UFRN.

{\small 

\end{document}